\documentclass[
aps,prl, amssymb, showpacs, twocolumn ]{revtex4}

\usepackage[american]{babel}
\usepackage{graphicx}
\usepackage{amsmath}
\usepackage{SIunits}

\begin{document}

\title{Exciton spin-flip rate in quantum dots determined by a modified local density of optical states}

\author{Jeppe Johansen}
\author{Brian Julsgaard}
\author{S\o ren Stobbe}
\author{J\o rn M. Hvam}
\author{Peter Lodahl}
\email{pelo@fotonik.dtu.dk.}

\affiliation{DTU Fotonik, Department of Photonics Engineering,
  Technical University of Denmark, \O rsteds Plads 343, DK-2800 Kgs. Lyngby,
  Denmark. http://www.fotonik.dtu.dk/quantumphotonics}

\begin{abstract}
The spin-flip rate that couples dark and bright excitons in
self-assembled quantum dots is obtained from time-resolved
spontaneous emission measurements in a modified local density of
optical states. Employing this technique, we can separate effects
due to non-radiative recombination and unambiguously record the
spin-flip rate. The dependence of the spin-flip rate on emission
energy is compared in detail to a recent model from the
literature, where the spin flip is due to the combined action of
short-range exchange interaction and acoustic phonons. We
furthermore observe a surprising enhancement of the spin-flip rate
close to a semiconductor-air interface, which illustrates the
important role of interfaces for quantum dot based nanophotonic
structures. Our work is an important step towards a full understanding of the complex dynamics of quantum dots in
nanophotonic structures, such as photonic crystals, and dark
excitons are potentially useful for long-lived coherent storage
applications.
\end{abstract}

\maketitle

 Understanding the optical
properties of solid-state quantum emitters is much required due to
their increasing importance for all-solid-state quantum photonic
devices for quantum information processing \cite{Lounis2005a}. Spin degrees of freedom of electrons and holes in semiconductor
quantum dots (QDs) impose exciton fine structure: The long-lived \emph{dark-exciton states} can
recombine after a phonon-mediated spin-flip process, whereby a
radiating bright exciton is formed \cite{Labeau2003a,Smith2005a}.
Dark exciton recombination is essential to explore: it influences
the performance of QD single-photon sources
\cite{Strauf2007a,Lund-Hansen2008a} and the quantum-optical
properties of QDs \cite{Reischle2008a}, and must be accounted for
when assessing inhibition of spontaneous emission in photonic
crystals \cite{Lodahl2004a,Julsgaard2008a}. Furthermore,
long-lived spin states in the solid state are required for spin
qubits \cite{Imamoglu1999a}, and dark excitons might be
attractive in this context since they can be prepared and read out
optically.

In a QD, the exciton spin lifetime is greatly extended compared to
a bulk medium or a quantum well due to the discrete energy
spectrum. The mechanism behind the spin-flip processes
has been debated in the literature
\cite{Roszak2007a,Tsitsishvili2005a}, and experimental tests for unbiased QDs have been lacking. Here we
employ controlled modifications of the local density of optical
states (LDOS) to determine the dark exciton spin-flip rate from
time-resolved spontaneous emission measurements. The LDOS is
modified by placing QDs at controlled distances from a
semiconductor-air interface, which was previously used to
distinguish radiative and non-radiative recombination processes of
bright excitons \cite{Johansen2008a}. Controlling the LDOS
provides a powerful tool to obtain detailed insight into QD
dynamics; thus cavity quantum electrodynamics was recently employed
for QD spectroscopy in the regime of continuous pumping
\cite{Winger08}. We observe a surprising dependence of the
spin-flip rate on the distance to the sample surface ranging
several hundred nanometers, and a sensitive dependence on the QD
emission energy. The latter is compared in detail to theory
describing the dependence of the spin-flip rate on QD size, and
 provides valuable insight on the exciton spin-flip mechanism in
 QDs.

\begin{figure}
  \center
  \includegraphics[width=\columnwidth]{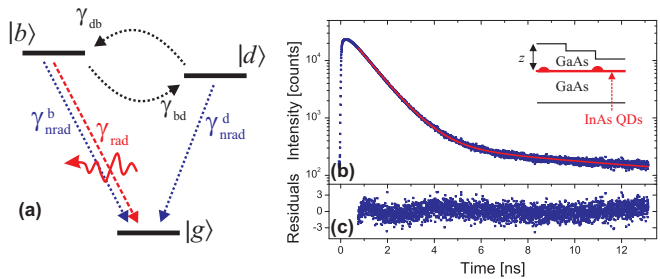}
  \caption{(Color online) \textbf{(a)}
    Three-level exciton scheme consisting of a bright $|b\rangle$ and a dark
    $|d\rangle$ state that are coupled through the spin-flip rates
    $\gamma_\mathrm{bd}$ and $\gamma_\mathrm{db}$. Radiative
    recombination ($\gamma_\mathrm{rad}$) to the ground state $|g\rangle$ (no exciton) is only possible for the
    bright state, while both the bright
    ($\gamma^\mathrm{b}_\mathrm{nrad}$) and the dark state
    ($\gamma^\mathrm{d}_\mathrm{nrad}$) can decay non-radiatively. \textbf{(b)} Typical decay curve acquired at a detection
    energy of $1.204 \, \mathrm{eV}$ (blue squares) at the sample with QDs positioned  $170\,\mathrm{nm}$ from the
    interface.
     The red line is a fit of the bi-exponential model to the data.
     \textbf{Inset} Sketch of the sample showing three different
     distances between the QDs and the GaAs-air interface.
     \textbf{(c)} The weighted residual obtained from the
     bi-exponential fit to the data resulting in $\chi^2_R=1.11$.}
\label{level-schemeANDdecay-curves}
\end{figure}

Figure ~\ref{level-schemeANDdecay-curves}(a) illustrates the fine
structure of the lowest exciton state for InAs/GaAs QDs \cite{Bayer2002a}. The exciton is formed from the
conduction band electron state (spin $1/2$) and the heavy-hole
valence band state (total angular momentum $3/2$). As a result,
four exciton states are formed. They are characterized by the
projections of the total angular momentum onto the growth axis,
which attain the values $\pm 1$ or $\pm 2$ for bright and dark
excitons, respectively. The splitting between dark and bright
energy levels $\Delta_\mathrm{db}$ is determined by the exchange
coupling between electron and hole spins, and the bright-exciton
levels are typically a few hundred $\micro \mathrm{eV}$ above the
dark-exciton levels. The bright excitons $|b\rangle$ decay to the
ground state $|g\rangle$ (no excitons) by emission of a photon
through a dipole-allowed transition with a rate
$\gamma_\mathrm{rad}.$ Radiative transitions from dark excitons
$|d\rangle$ are forbidden, however, they can decay through a
phonon mediated spin-flip process (rate $\gamma_\mathrm{db}$)
transforming the dark exciton into a bright exciton. The rate of
the reverse process is denoted $\gamma_\mathrm{bd}.$ Finally
non-radiative recombination was recently proven to be significant
for self-assembled QDs \cite{Johansen2008a}, i.e., bright and dark excitons can recombine
non-radiatively with rates denoted $\gamma_\mathrm{nrad}^b$ and
$\gamma_\mathrm{nrad}^d$, respectively.

We have measured spontaneous emission decay dynamics of InAs QDs
positioned at $28$ different distances from a GaAs/air interface,
see inset of Fig.~\ref{level-schemeANDdecay-curves}(b). Further
experimental details can be found in Refs.
\cite{Johansen2008a,Stobbe2009a}. We record the number of photons emitted per
time interval $N(t)$ from the QDs. Solving the rate equations for
the three-level system of Fig.~\ref{level-schemeANDdecay-curves}(a)
results in a bi-exponential decay $N(t) =
A_\mathrm{f}e^{-\gamma_\mathrm{f}
t}+A_\mathrm{s}e^{-\gamma_\mathrm{s} t}.$ Here subscripts s and f
denote slow and fast decay rates, respectively.
Figure~\ref{level-schemeANDdecay-curves}(b) shows an example of a
spontaneous emission decay curve together with a bi-exponential
fit obtained with a fixed background level to account for the
measured contribution from dark-counts and after-pulsing of the
detector. Excellent agreement between the experiment and the model
is apparent from the weighted residuals reproduced in
Fig.~\ref{level-schemeANDdecay-curves}(c).

The four parameters $\gamma_\mathrm{f/s}, A_\mathrm{f/s}$ are obtained by fitting the experimental decay curves. We have
$\gamma_\mathrm{f} \simeq
\gamma_\mathrm{rad}+\gamma_\mathrm{nrad}^\mathrm{b}$ and
$\gamma_\mathrm{s} \simeq \gamma_\mathrm{nrad}^\mathrm{d},$ where
it has been assumed that the spin-flip rate is slow compared to
both radiative and non-radiative processes, which is a very good
approximation as seen below. Thus, the spin-flip rate cannot be obtained
from the measured rates $\gamma_\mathrm{f/s}$ since they are
dominated by other decay contributions. However, the amplitudes
$A_\mathrm{f/s}$ contain additional information and the
dark-bright spin-flip rate is contained in their ratio
\begin{eqnarray}\label{eq:RatioAmplitudes}
  \frac{A_\mathrm{f}(z)}{A_\mathrm{s}(z)}=\frac{\gamma_\mathrm{f}(z)-\gamma_\mathrm{s}}{\gamma_\mathrm{db}(z)}
  \frac{\rho_\mathrm{b}(t=0)}{\rho_\mathrm{d}(t=0)}-1,
\end{eqnarray}
where $\rho_\mathrm{b}(t=0)/\rho_\mathrm{d}(t=0)$ is the ratio
between the initial populations of bright and dark excitons after
an excitation pulse and $z$ is the distance to the interface. The
repetitive nature of the experiment implies that the slow
amplitude contains contributions
from previous excitation pulses, which is accounted for by
correcting the amplitudes according to
$A_{\mathrm{s}} \rightarrow A_{\mathrm{s}}
\left(1-e^{-\gamma_\mathrm{s} T} \right)$ where $T=13.5 \:
\mathrm{ns}$ is the repetition period of the excitation laser. The
spin-flip rate is obtained from experimentally determined
parameters by:
\begin{eqnarray}\label{eq:SpinFlipRate}
  \gamma_\mathrm{db}(z)=\frac{\gamma_\mathrm{f}(z)-\gamma_\mathrm{s}}{1+A_\mathrm{f}(z)/A_\mathrm{s}(z) }
  \frac{\rho_\mathrm{b}(t=0)}{\rho_\mathrm{d}(t=0)}.
\end{eqnarray}
It is essential that the ratio of the amplitudes enters in
Eq.~(\ref{eq:SpinFlipRate}) since the absolute size of
$A_\mathrm{f/s}$ will depend on, e.g., the total collection
efficiency of the radiated light or the number of QDs probed.
Since the bright and dark excitons recombine radiatively on the
same transition (cf. Fig.~\ref{level-schemeANDdecay-curves}(a))
these dependencies do not contribute to the amplitude ratio. Quite
remarkably, assessing the amplitudes allows for extraction of spin-flip
rates that are slower than the repetition period of the
measurement.

\begin{figure}[ht]
  \center
  \includegraphics[width=\columnwidth]{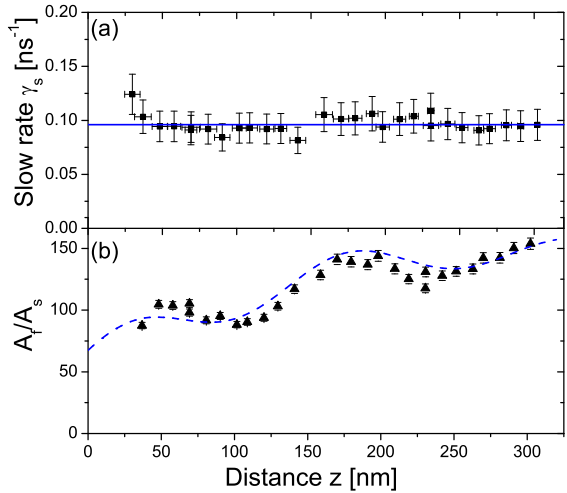}
  \caption{(Color online)
     \textbf{(a)} The slow decay rate as a function of distance to the
   interface measured at $1.204 \: \mathrm{eV}$. No systematic dependence on distance is observed confirming that
   the slow rate is due to non-radiative decay of dark excitons.
   The blue line indicates the averaged rate.
     \textbf{(b)} Measured ratio of the fast and slow amplitudes versus
     distance. The dashed blue curve is obtained by multiplying the
     calculated LDOS with an overall exponential decrease of the spin-flip rate with the distance to the interface.}
\label{fig:SlowComponent_nonrad_13ns_b}
\end{figure}

In Fig.~\ref{fig:SlowComponent_nonrad_13ns_b} the slow decay rate
and the ratio of the amplitudes are plotted versus distance to the
interface. The fast rate (not shown) varies periodically with
distance in full agreement with the modified LDOS caused by the
reflecting interface \cite{Johansen2008a}. In contrast, the slow
decay rate is constant within the error-bars of the measurement,
which confirms that it is dominated by non-radiative recombination
of dark excitons $(\gamma_\mathrm{nrad}^\mathrm{d})$. The
average decay rate of
 $\gamma_\mathrm{s} = 0.097 \pm 0.008 \:
\mathrm{ns}^{-1}$ matches very well with the non-radiative decay
rate of bright excitons at the same emission energy, which is expected since
the bright and dark exciton binding energies are very close. $A_\mathrm{f}(z)/A_\mathrm{s}(z)$
is expected to vary proportional to the LDOS through
$\gamma_\mathrm{f}(z)$ (cf. Eq.~(\ref{eq:RatioAmplitudes})) and
indeed clear oscillations with a period matching the LDOS is
observed. In addition the spin-flip rate is found to vary, which
gives rise to the overall increase of the amplitude ratio with
$z$. Assuming an exponential decrease of the spin-flip rate with
$z$, which will be discussed further below, and multiplying by the
calculated LDOS we can model the experimental data very well, see
Fig.~\ref{fig:SlowComponent_nonrad_13ns_b}(b).

In order to extract the spin-flip rate from
Eq.~(\ref{eq:SpinFlipRate}), we need to estimate the ratio of the
initial populations of bright and dark excitons. Due to the
non-resonant excitation, the feeding of the QDs is non-geminate
\cite{Baylac1995}, thus dark and bright excitons are formed with
equal probability.
A slight imbalance between the initial populations is
nonetheless created due to the finite probability of
creating biexcitons. The recombination of biexcitons is dominated by radiative decay, which always leaves behind a bright
exciton and thus increase $\rho_\mathrm{b}(0)$ relative to
$\rho_\mathrm{d}(0)$. The long lifetime of the dark excitons
moreover increases the probability of biexciton formation since a
residual population of dark excitons is persistent when the
subsequent excitation re-excites the QDs.
In the present experiment the excitation density was fixed such
that on average $0.1$ excitons were generated per QD and by
solving the rate equations for the QD population including the
biexciton level we find
${\rho_\mathrm{b}(t=0)}/{\rho_\mathrm{d}(t=0)} \approx 1.25$.

\begin{figure}
  \center
  \includegraphics[width=\columnwidth]{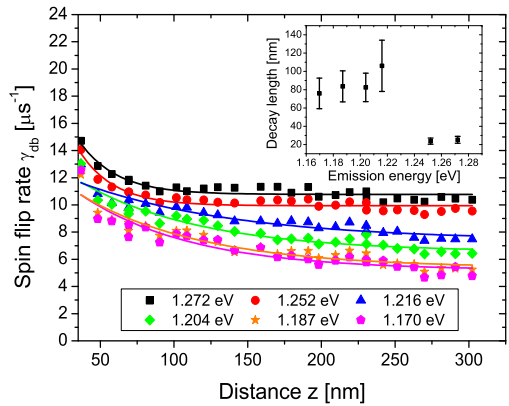}
  \caption{(Color online)
    The spin-flip rates versus distance $z$ to the interface for six different emission energies.
    The plotted curves are fits to the experimental data assuming an exponential decay of the spin-flip rate with the distance to the interface. The inset shows the exponential decay length versus emission energy. }
\label{fig:SpinFlipDistance}
\end{figure}

The spin-flip rate versus distance to the interface is presented
in Fig.~\ref{fig:SpinFlipDistance} for six different emission
energies. It is found to vary between $4.7\:
\mathrm{\micro s}^{-1}$ and $14.7 \: \mathrm{\micro s}^{-1}$
corresponding to long-lived exciton spin lifetimes between $68\:
\mathrm{ns}$ and $215\: \mathrm{ns}.$ Surprisingly the spin-flip
rate is observed to depend exponentially on distance to the
interface, which is observed for all emission energies.
 The characteristic decay length varies between $24 \: \mathrm{nm}$ and
$106 \: \mathrm{nm}.$ This increase of the spin-flip rate in
the vicinity of the interface could be caused by enhancement of acoustic phonons at the interface. In this case, we estimate the phonon wavelength
 from $\lambda_{\mathrm{ph}} = h v/\Delta_{\mathrm{db}}$ where $h$ is Planck's constant, $v$ is the sound speed and $\Delta_{\mathrm{db}}$ the exchange energy splitting between bright and dark states that must be matched by the phonons. For GaAs we estimate $\lambda_{\mathrm{ph}} \sim 65-85 \: \mathrm{nm}$ for longitudinal phonons using the range of exchange energies relevant in the experiment, which matches the length scale observed in the experiment.

The above results illustrate clearly the importance of nearby
surfaces when seeking a quantitative understanding of the dynamics
of QDs in nanostructures. Recent examples of intricate surface
effects include the increased emission rate observed near a
semiconductor interface \cite{Stobbe2009a}, while charge
trapping near surfaces was suggested as the mechanism responsible
for the surprisingly large QD-nanocavity coupling efficiency at
very large detunings observed under non-resonant excitation
\cite{Hennessy2007a,Kaniber2008a,Laucht2009a}. These are examples
of the new physics found in solid-state
implementations of quantum electrodynamics experiments.

\begin{figure}
  \center
  \includegraphics[width=\columnwidth]{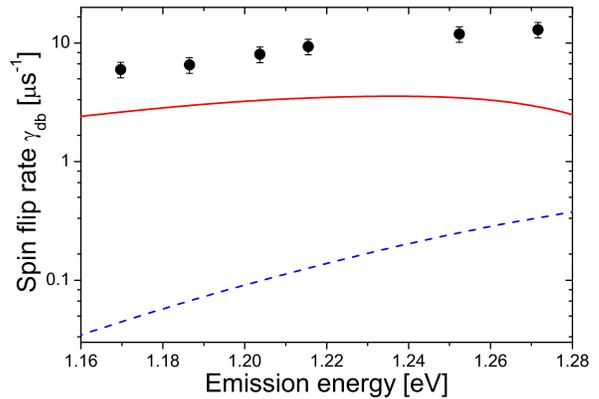}
  \caption{(Color online) Measured spin-flip rate at different emission energies for the sample with $(z=302\:\mathrm{nm})$ (solid black circles). The theoretically predicted spin-flip rate using the model of Ref. \cite{Roszak2007a} for parameters that reproduce the energy dependence of the radiative decay rate (dashed blue line) and for optimized parameters (solid red line).}
\label{fig:SpinFlipEnergy}
\end{figure}

Let us now consider the energy dependence of the spin-flip rate,
which allows to assess how the spin-flip process varies with QD size.
In order to minimize the effect of the interface, we plot in
Fig.~\ref{fig:SpinFlipEnergy} the spin-flip rate versus energy for
the sample with largest QD-interface distance
$(z=302\:\mathrm{nm}).$ The spin-flip rate is observed to increase
significantly with the emission energy, and varies from
$6\:\mathrm{ns}^{-1}$ at $1.170\:\mathrm{eV}$ to
$13\:\mathrm{ns}^{-1}$ at $1.272\:\mathrm{eV}$ meaning that the
characteristic spin-flip time can be prolonged by more than a
factor of 2 by varying the QD size.

The energy-dependence of the spin-flip rate can be compared to
theory using the model introduced in Ref. \cite{Roszak2007a}. In
this work the dark-bright exciton spin-flip rate was
calculated due to the combined effect of short-range exchange
interaction and acoustic phonons. A complex behavior is expected
since the spin-flip rate is predicted to depend on the energy
splitting between the lowest and first excited exciton states
$\Delta_{12}$, the exchange energy splitting $\Delta_{\mathrm{db}}$,
as well as the size of the electron and hole wavefunctions.

The experimental data for the
energy-dependent spin-flip rates are compared to theory using
experimentally realistic parameters: By recording the emission
spectrum of the QDs at high pump power where the ground state is
saturated, we obtain $\Delta_{12} = 120 \pm 20 \: \mathrm{meV}.$
The exchange splitting depends strongly on the indium mole fraction ~\cite{Fu1999} and it is enhanced in QDs due to the strong localization of the electron and hole wavefunctions as described by the enhancement factor~\cite{Fishman1994a}. Furthermore, the spin-flip rate depends on the electron and hole wavefunctions through a form factor~\cite{Grodecka2005}. We have calculated the electron and hole wavefunctions using the theory and parameters of Ref.~\cite{Stobbe2009a}. All other parameters are taken from Ref.~\cite{Roszak2007a} and the references given above.

The comparison to theory is presented in Fig.~\ref{fig:SpinFlipEnergy}. The dashed blue curve is for parameters that reproduce the frequency dependence of the radiative decay rate~\cite{Stobbe2009a} (aspect ratio: $1/2$, indium mole fraction: $0.46$, QD heights: between 4~nm and 5.8~nm). The observed increase of the spin-flip rate with energy is reproduced by the theory. However, the calculated rates are two orders of magnitude smaller than the measured values. We have explored the origin of this deviation by optimizing the model parameters in order to maximize the spin-flip rate, see the solid red curve in Fig.~\ref{fig:SpinFlipEnergy} (aspect ratio: $1/3$, indium mole fraction: $0.90$, QD heights: between 1.9~nm and 2.4~nm). In this case the predicted spin-flip rates approach the measured rates, although a systematic discrepancy is still observed. Furthermore, a rather weak frequency dependence of the spin-flip rate is predicted, since the finite confinement potentials impose lower limits to the achievable compression of the wavefunctions. We conclude that additional spin-flip processes must be included in a complete theory. Additional contributions may arise due to spin-orbit coupling \cite{Tsitsishvili2005a} or long-range exchange effects \cite{Luo09}.

In conclusion, we have measured the rate for spin flip from dark
to bright excitons using time-resolved fluorescence spectroscopy.
The spin-flip rate increases significantly when approaching the
sample surface suggesting an enhancement due to surface acoustic
phonons. The energy dependence of the spin-flip rate was compared to a recent
theory, where the spin flip is induced by short-range
exchange interaction between electrons and holes while the
required energy is provided by acoustic phonons. Our results
illustrate the importance of taking the QD fine structure into
account when interpreting luminescence experiments, and will be
important in order to obtain quantitative understanding of
light-matter interaction in QD based optical devices.

We gratefully acknowledge the Danish Research Agency for financial
support (projects FNU 272-05-0083, 272-06-0138 and FTP
274-07-0459).

\end{document}